\documentclass[14pt]{extarticle}
\usepackage[utf8]{inputenc}
\usepackage{mathtools}
\usepackage{amsfonts}
\usepackage{amsmath}
\usepackage{amssymb}
\usepackage{youngtab} 
\usepackage{fancyhdr}
\usepackage{hyperref}

\newcommand\be{\begin{equation}}
\newcommand\ba{\begin{eqnarray}}
\newcommand\ee{\end{equation}}
\newcommand\ea{\end{eqnarray}}
\numberwithin{equation}{section}

\usepackage[
   citestyle=numeric-comp,
   sorting=none,
   maxnames=10,
  ]{biblatex}
 \renewbibmacro{in:}{}  

\DeclareFieldFormat{postnote}{#1}
\DeclareFieldFormat{multipostnote}{#1} 
\DeclareFieldFormat{pages}{#1}

\title{Level-rank duality for a Universal Topological Quantum Computer}

\author{Howard J. Schnitzer\footnote{schnitzr@brandeis.edu} \\ Department of Physics \\ Brandeis University \\ Waltham, MA 02454}

\begin{document}
\flushbottom
\maketitle

\thispagestyle{fancy}
\renewcommand{\headrulewidth}{0pt}
\rhead{BRX-TH 6635}

\begin{abstract}
It is shown, using level-rank duality that a universal topological quantum computer based on Chern-Simons theory for $\mathrm{SU}(2)_3$ also implies an analogous universal quantum computer based on $\mathrm{SU}(3)_2$. Suggestions are made for the possible role of level-rank duality in entanglement from topology.
\end{abstract}

\clearpage

\section{Introduction}

Quantum computing and quantum information theory have become a focus of various strategies for investigations of topological field theories and entanglement. In particular, Chern-Simons theories \cite{Witten1988,Witten1989} play a role in topological quantum computers \cite{FreedmanLarsenWang2002a,FreedmanLarsenWang2002b,FreedmanKitaevEtAl2003,FreedmanKitaevWang2002,FreedmanNayakWalker2006,FreedmanNayakWalker2005,KitaevPreskill2006,SaltonSwingleWalter2017,NayakSimonEtAl2008,MelnikovMironovEtAl2018}, as well as entanglement in topological field theories. One feature of Chern-Simons theories \cite{NaculichSchnitzer1990a,NaculichSchnitzer1990b,NaculichRiggsSchnitzer1990,MlawerNaculichEtAl1991,NaculichRiggsSchnitzer1993} that has not received adequate attention is the impact of level-rank duality \cite{NaculichSchnitzer1990a,NaculichRiggsSchnitzer1990,NaculichRiggsSchnitzer1993} on these topics. One example of a topological quantum computer is considered in 
\cite{FreedmanLarsenWang2002a}, where one presents topological modular functors which are universal. In \cite{FreedmanLarsenWang2002a} this is analysed for $\mathrm{SU}(2)_3$ Chern-Simons theory and extended in \cite{FreedmanLarsenWang2002b} to $\mathrm{SU}(2)_K$ for $K \neq 1,2$ or $4$. In this paper we apply level-rank duality to the results of \cite{FreedmanLarsenWang2002a} to obtain a universal quantum computer for $\mathrm{SU}(3)_2$. Thus, this relates a quantum computer for qubits to one with qutrits. Since our discussion is a corollary of \cite{FreedmanLarsenWang2002a}, we just concentrate on the role of level-rank duality in the analysis.

In Sec. \ref{sec:Level-rank duality} we review essential features of level-rank duality so as to make our discussion reasonably self-contained, with more details to be found in the original references. Section \ref{sec:braid group} presents the level-rank duality of the Jones representation of the braid group. Section \ref{sec:SU(3)_2} applies the level-rank duality $\mathrm{SU}(2)_3 \leftrightarrow \mathrm{SU}(3)_2$ to describe universal quantum computation for $\mathrm{SU}(3)_2$ Chern-Simons theory. Section \ref{sec:discussion} presents suggestions for applications of level-rank duality related to entanglement in topological field theories \cite{LarsenWang2005,BondersonFidkowskiEtAl2013,BarkeshliJianQi2013,BondersonFidkowskiEtAl2016,BarkeshliFreedman2016,DongFradkinEtAl2008,BalasubramanianFlissEtAl2017,BalasubramanianDeCrossEtAl2018,ChunBao2017}.

\section{Level-rank duality} \label{sec:Level-rank duality}
Level-rank duality was discovered for WZW theories in \cite{NaculichSchnitzer1990a,NaculichSchnitzer1990b}, and further developed in \cite{NaculichRiggsSchnitzer1990,MlawerNaculichEtAl1991,NaculichRiggsSchnitzer1993}. In \cite{NaculichRiggsSchnitzer1990} the level-rank duality of quantum dimensions and the eigenvalues of braid matrices is discussed for $\mathrm{SU}(N)_K$, $\mathrm{Sp}(N)_K$ and tensor representations of $\mathrm{SO}(N)_K$. Further results for the role of level-rank dualities for braid matrices can be found in \cite{MlawerNaculichEtAl1991,NaculichRiggsSchnitzer1993}.

Level-rank duality of WZW and Chern-Simons theories involve maps of various quantities of affine $\mathrm{G}(N)_K$ to those of $\mathrm{G}(K)_N$, as exemplified by the relationship of $\mathrm{SU}(N)_K$ to $\mathrm{SU}(K)_N$. A common misconception is that the duality is a map of a Young tableau to a transposed tableau. Rather it is a map between equivalence classes. For example $\mathrm{SU}(2)_3$ has four irreducible representations while $\mathrm{SU}(3)_2$ has six. More generally irreducible representations of $\mathrm{SU}(N)_K$ are described by reduced Young tableaux, which are not 1-to-1 with those of $\mathrm{SU}(K)_N$.

For $\mathrm{SU}(N)_K$, define
\be q = \exp [ 2\pi i / (N+K)] \ee
and the quantum dimension of $X$, with 
\be [X] = (q^{X/2} - q^{-X/2})/(q^{1/2}- q^{-1/2})  \label{eqn:quantum dimension} \ee
Therefore $q$ and $[N] = [K]$ are level-rank invariant. In this paper we will be concerned with the relationship of $\mathrm{SU}(3)_2$ to $\mathrm{SU}(2)_3$.

Details for the map of representations of $\mathrm{SU}(N)_K$ to those of $\mathrm{SU}(K)_N$ are described in \cite{MlawerNaculichEtAl1991}. Also important for our considerations is the fusion of irreducible representations $a \times b = \sum c$ of $\mathrm{SU}(N)_K$, which is also presented in detail in \cite{MlawerNaculichEtAl1991}, where $a$ and $b$ are described by \emph{reduced} Young diagrams. In general the number of boxes of $c$ may be less than the sum of those of $a$ and $b$ because the representation of $c$ also is reduced. If $\tilde c$ is the transpose of the tableau of $c$, then $c$ is dual to a representation $\sigma^\Delta (\tilde c)$ which is cominimally equivalent to $\tilde c$. This construction is described in  \cite{MlawerNaculichEtAl1991}. These tools then allow us to apply level-rank duality to the Jones representation of the braid group, and construct a $\mathrm{SU}(3)_2$ universal quantum computer from that for $\mathrm{SU}(2)_3$.

\section{Level-rank duality of the Braid Group} \label{sec:braid group}

Our discussion of the level-rank dual of the Jones representation is based on Sec. 3 of \cite{FreedmanLarsenWang2002a}. Begin with the representation of the Temperley-Lieb-Jones algebras $A_{\beta,n}$ generated by 1 and the projections $e_1, \ldots, e_{r-1}$, for some integer $r \geq 3$, 
with
\begin{equation}
 q= \exp\left( \frac{2\pi i}{r} \right)
\end{equation}
and

\begin{align}
 \nonumber  \qquad & e_i^2 =e_i; \quad e_i^* = e_i
 \\
 \qquad & e_i e_{i+1} e_i = \beta^{-1} e_i
  \\
 \nonumber  \qquad & e_i e_j = e_j e_i \quad \text{if} \quad |i-j| \geq 2
\end{align}
where
\begin{equation}
 \beta = [2]^2 = q + \bar{q} + 2
\end{equation}
with the quantum dimension defined by \eqref{eqn:quantum dimension}. 

The Jones representation of $A_{\beta,n}$ has the feature that it splits into a direct sum of irreducible representations indexed by a 2-row Young diagram. A Young diagram is labeled by $\lambda = [\lambda_1, \ldots, \lambda_s], \lambda_1 \geq \lambda_2 \geq \ldots \geq \lambda_s$. If 
$s \leq 2$
and $\lambda_1 - \lambda_2 \leq r - 2$, 
then the diagram has at most 2 rows. The representation $A_{\beta,n}$ is a direct sum of irreducible representations $\pi_\lambda^{(2,r)}$ over all $(2,r)$ Young diagrams $\lambda$. For $\mathrm{SU}(2)_K$ the diagram is not reduced, in the sense of Sec. \ref{sec:Level-rank duality}, with $K= r-2$, if the diagram has two rows. For $\mathrm{SU}(2)_3, r=5,$ we consider Chern-Simons theory at the 5\textsuperscript{th} root of unity, as discussed in 
\cite{FreedmanLarsenWang2002a}.

Following Sec. 3 of \cite{FreedmanLarsenWang2002a}, the Jones representation of the braid group is defined by
\begin{equation}
 \rho_{\beta,n}(\sigma_i) = q - (1+q)e_i.
\end{equation}
It follows that the representation $\pi_\lambda^{(2,r)}$ is a matrix with only $2 \times 2$ blocks and $1\times 1$ blocks whose entries are either $0$ or $1$. The $2 \times 2$ blocks are
\begin{equation} \label{eqn:2x2block}
 \begin{pmatrix}
  \alpha_{t,i} & \beta_{t,i} \\
  \beta_{t,i} & 1 - \alpha_{t,i}
 \end{pmatrix},
\end{equation}
with \eqref{eqn:2x2block} a projection. Then from \cite{FreedmanLarsenWang2002a} the Jones representation of the braid group denoted by $\rho_{\beta,n}$ is 
\begin{equation} \label{eqn:P_beta,n}
 \rho_{\beta,n} = B_n \rightarrow A_{\beta,n} \rightarrow U(N_{\beta,n}).
\end{equation}
When $|q| = 1$, the representation $\rho_{\beta,n}$ is unitary, and splits into a direct sum of representations of $(2,r)$ Young diagrams, where a sector of a particular Young diagram is denoted by $\rho_{\lambda,\beta,n}$. One identifies sectors of the Jones representation with representations of braid groups for Chern-Simons theory $\mathrm{SU}(2)_3$, where here $r=5, K=r-2,$ and $q=\exp \frac{2\pi i}{r}$. Since the representations $\rho_{\lambda,\beta,n}$ are functions of $q$ only, it follows that the Jones representations for $\mathrm{SU}(2)_3$ are level-rank dual. That is, the Jones representations of $\mathrm{SU}(3)_2$ are identical to those of $\mathrm{SU}(2)_3$.

Equation \eqref{eqn:P_beta,n} is generalized in Theorem 0.1 of \cite{FreedmanLarsenWang2002b} for $r \geq 5$,
\begin{equation}
 r\neq 6, 10, \quad n \geq 3 \quad \text{or} \quad r=10, \quad n \geq 5
\end{equation}
for the $n$-strand braid group $B_n$, and where
\begin{equation}
 \begin{aligned}
  \rho_n^{2,r} = \bigoplus_{\lambda \in \Lambda_n^{(2,r)}} \rho_\lambda^{(2,r)}
  \\
  B_n \rightarrow \prod_{\lambda \in \Lambda_n^{(2,r)}} U(\lambda)
 \end{aligned}
\end{equation}
is a unitary representation of $B_n$.

\section{$\mathrm{SU}(3)_2$ as a universal quantum computer} \label{sec:SU(3)_2}
Section 2 of \cite{FreedmanLarsenWang2002a} discusses $\mathrm{SU}(2)_3$ Chern-Simons theory as a universal quantum computer. In this section we argue, using level-rank duality, that $\mathrm{SU(3)}_2$ Chern-Simons theory is also a universal quantum computer. Our discussion closely follows that of \cite{FreedmanLarsenWang2002a}. This allows us to just emphasize the role of level-rank duality in the discussion, referring the reader to \cite{FreedmanLarsenWang2002a} for further details.

The state space of $\mathrm{SU}(3)_2$ for the quantum computer is
\begin{equation}
S_k = (\mathbb{C}^3)^k    
\end{equation}
which consists of $k$ qutrits i.e, the disjoint union of $k$-copies of the basic $3$-level system, $\mathbb{C}^3 =span(|0 \rangle, |1 \rangle, |2 \rangle)$. For each integer $k$, choose an inclusion 
\begin{equation}
 \begin{split}
    S_k & \underset{i}{\rightarrow} \widetilde{V} (D^2, 3k \quad \text{marked points}) \\
        & = \widetilde{V} (D^2,3k) 
 \end{split}    
\end{equation}

The irreducible representations of $\mathrm{SU(3)}_2$ are labelled by six reduced Young diagrams
\begin{equation}
    \begin{split}
    &\{(.); {\tiny \yng(1)}; {\tiny \yng(2)}; {\tiny \yng(1,1)}; {\tiny \yng(2,1)}; {\tiny \yng(2,2)} \} \\
    & \equiv \{0,1,2,3,4,5\}
    \end{split}
\end{equation}

Let $D$ be the unit 2-dimensional disk, with a subset of $3k$ marked points on the x-axis. Picture $k$ disjoint sub-disks, $D_i ; \quad 1 \leq i \leq j $ each containing 3 marked points in the interior (which serve as qutrits). The $ \binom{k}{2}$ disks \\
$D_{i,j}; \textrm{with} \,  l \neq i \, \mathrm{or} \, j, \quad 1 \leq i \leq j \leq k$ contain $D_i$ and $D_j$ , but with $D_{i,j} \cap D_{l} = \phi$.

Define $\widetilde{V}_k$ the $\mathrm{SU}(3)$ Hilbert space of $k$ marked points in the interior with labels $(i)= {\tiny \yng(1)}$, and $l$ on $\partial D$. The various ways that $\widetilde{V}^k_l$ occurs arises from the gluing axioms of $\mathrm{SU}(3)_2$. [See Mlawer et.al ref \cite{MlawerNaculichEtAl1991} for the explicit multiplication rules]. One finds by direct computation 
\begin{equation} \label{su3dim}
    \begin{split}
       & \dim \widetilde{V}_3^{(0)} =2 ; \quad \dim \widetilde{V}_3^{[2,1]} =1 \\
       & \dim \widetilde{V}_6^{[2,1]}=8 ; \quad \dim \widetilde{V}_6^{(0)} = 5
    \end{split}
\end{equation}
These are numerically identical to the level-rank duals of those of $\mathrm{SU}(2)_3$, i.e,
\begin{equation} \label{su2dim}
    \begin{split}
       & \dim V_3^{1} =2 ; \quad \dim V_3^{3} =1 \\
       & \dim V_6^{2}=8 ; \quad \dim V_6^{0} = 5
    \end{split}
\end{equation}

The dimensions of $\widetilde{V}_{k,l}$ in (\ref{su3dim}) are obtained from direct computation as well as from the level-rank dual maps of (\ref{su2dim}), using the duality rules of ref \cite{MlawerNaculichEtAl1991}. Thus (\ref{su3dim}) and (\ref{su2dim}), together with level-rank duality means that the Jones representations $\rho_{\lambda}^{(2,5)}$ of the braid group are identical for $\mathrm{SU}(2)_3$ and $\mathrm{SU}(3)_2$. [See \cite{FreedmanLarsenWang2002a} for explicit matrix representations  for (2,5) Young diagrams]. 
Identify $\widetilde{V} (D_i , \text{with 3 points, boundary label 1})$ $ \equiv \widetilde{V}  \equiv  \mathbb{C}^3$ as the fundamental qutrit. Repeating the discussion of equations 5 and 4 of \cite{FreedmanLarsenWang2002a}, now for $\mathrm{SU(3)}_2$, the action of $\mathrm{B}(3)$ on $D_i$ yields 1-qutrit gates and $\mathrm{B}(6)$ on $D_{i,j}$ gives the action of 2-qutrit gates. One can evolve a $2$-qutrit gate $g$ acting unitarily on $\mathbb{C}_i^3 \otimes \mathbb{C}_j^3$ and as (identity) on $\mathbb{C}_l^3, \quad l \neq i \, \textrm{or} \, j$. However, one cannot control the phase of the output, so that those outputs will only be approximations of the desired gate $g$. The approximation issues are addressed by Theorem (2.1) of \cite{FreedmanLarsenWang2002a}. The phase issue $v \rightarrow v^*$ is harmless, since one only measures qutrits. The significant approximation is the qutrit analogue of Theorem (2.1) of \cite{FreedmanLarsenWang2002a}. For $\mathbb{C}_i^3 \otimes \mathbb{C}_j^3 \rightarrow \mathbb{C}_i^3 \otimes \mathbb{C}_j^3$ there is a braid $b_l$ of length $\leq l$ in generators $\sigma_i$ and $\sigma_i', \quad 1 \leq i \leq n-1$, so that 
\begin{equation}
    l \leq C \Big(\frac{1}{\epsilon} \Big)^2 
\end{equation}

The main work in proving the analogue of Theorem (2.1) is that the closure of the image of the representation 
$\rho : \mathrm{B}(6) \rightarrow \mathrm{U}(5) \times \mathrm{U}(8)$ contains $\mathrm{SU}(5) \times \mathrm{SU}(8)$. This requires the analogue of the density theorem (4.1) of \cite{FreedmanLarsenWang2002a}, which is in fact level-rank dual. Let $\rho = \rho_{[3,3]} \oplus \rho_{[4,2]}$ and $\mathrm{B}(6) \rightarrow \mathrm{U}(5) \times \mathrm{U}(8)$ be the Jones representation of $\mathrm{B}_6$ at the fifth root of unity, i.e, $q=e^{\frac{2 \pi i}{5}}$. Then the closure of the image of $\rho(\mathrm{B}(6)$ in $U(5) \times \mathrm{U}(8)$ contains $\mathrm{SU}(5) \times \mathrm{SU}(8)$. Since the Jones representation is level-rank dual, theorem (2.1) of \cite{FreedmanLarsenWang2002a} applies to $\mathrm{SU}(3)_2$.

Therefore, exactly as in \cite{FreedmanLarsenWang2002a} the computational model based on Chern-Simons theory for $q=e^{\frac{2 \pi i}{5}}$ is well defined for both $\mathrm{SU}(2)_3$ and $\mathrm{SU}(3)_2$ as level-rank dual pairs, and are polynomial equivalent to the quantum circuit model, as efficiently approximated by an intertwining action of a braid on this functor state space.

\section{Discussion} \label{sec:discussion}
In the body of the paper we used level-rank duality of $\mathrm{SU}(N)_K$ and the analysis of \cite{FreedmanLarsenWang2002a} to argue that the computational model based on Chern-Simons theory for $\mathrm{SU}(2)_3$ is universal for quantum computation implies the same for Chern-Simons theory for $\mathrm{SU}(3)_2$. From theorem (0.1) of \cite{FreedmanLarsenWang2002b} one conjectures that $\mathrm{SU}(2)$ Chern-Simons theory for $\mathrm{SU}(2)_K$ is universal for quantum computation $K \neq 1,2,$ or $4$. Then level-rank duality would suggest that Chern-Simons theory for $\mathrm{SU}(K)_2$, $K \neq 2$ or $4$ also provides a universal model for quantum computation. 

One may ask whether Chern-Simons theory for $\mathrm{SO(N)}_K$ or $\mathrm{Sp(N)}_K$ can provide a model for a universal quantum computation ? For these cases the level-rank duality of the quantum dimensions is \cite{NaculichRiggsSchnitzer1990}
\be [N-1] +1 = [K-1] + 1 \ee
for non-spinor representation of $\mathrm{SO}(N)_K$ and 
\be [N + 1/2] -1 = [K+ 1/2] \ee
for $\mathrm{Sp}(N)_K$.

Representation of the braid group $B_n$ for $\mathrm{SO}(N)_K$ and $\mathrm{Sp}(N)_K$ are discussed in eqns (20) and (21) of ref. \cite{NaculichRiggsSchnitzer1990}, eqns (2.8), (2.9), (4.2), (4.6), (4.7) and (6.19) – (6.23) of ref. \cite{NaculichRiggsSchnitzer1993}. We conjecture that there are examples of Chern-Simons theories for $\mathrm{Sp}(N)_N$ and tensor representations of $\mathrm{SO}(N)_K$ which can provide a model for a universal quantum computer, based on representations of the Hecke algebra for these theories. If so, level-rank duality should play a role for these cases. 

Larsen and Wang \cite{LarsenWang2005} considered a $(2+1)$ dimensional topological field theory (TQFT) on a closed oriented surface of genus $g$; in particular the $\mathrm{SO}(3)$ TQFT at an $r$-th root of unity, $r \geq 5$ prime. Their proof depends on the result that a Dehn twist acts with $(r-1)/2$ distinct eigenvalues. For $r \geq 7$, the two-eigenvalue property no longer holds. Using this result \cite{SaltonSwingleWalter2017}, their theorem 2, proves that $\mathrm{SO}(3)_K$ Chern-Simons theory at level $K \geq 8$, with $K-3$ an odd prime can approximate any state in the n-torus Hilbert space to an arbitrary precision by path integration, which can be placed in the context of topological quantum computation. However \cite{SaltonSwingleWalter2017} deals with “state universality” which is distinct from universality of quantum computers discussed in the body of this paper. Schemes which are not universal for topological quantum computation can be made universal by incorporating certain topology changing equations \cite{FreedmanNayakWalker2006,FreedmanNayakWalker2005,BondersonFidkowskiEtAl2013,BarkeshliJianQi2013,BondersonFidkowskiEtAl2016,BarkeshliFreedman2016}. Both ideas of universality are closely related to properties of the mapping class group. 
Level-rank duality could then be applicable to theorem 2 of \cite{SaltonSwingleWalter2017} relating tensor representations of $\mathrm{SO}(3)_K$, to $\mathrm{SO}(K)_3$, where $K \geq 8$, and $K-3$ an odd prime. 

Also related to these issues \cite{DongFradkinEtAl2008,BalasubramanianFlissEtAl2017,BalasubramanianDeCrossEtAl2018,ChunBao2017,DwivediSinghEtAl2017} is entanglement from topology in Chern-Simons theories. Many of these examples should also be explored from the point of view of level-rank duality.

\section*{Acknowledgements}

This research was supported in part by the DOE by grant DE-SC000987, we thank Brian Swingle for conversations and Isaac Cohen, Alastair Grant-Stuart, Harsha Hampapura, and Andrew Rolph for assistance in the preparation of the paper.

\printbibliography


\end{document}